# Optimization of the collimation system for CSNS/RCS with the robust conjugate direction search algorithm*

Hong-Fei Ji (纪红飞)[1,2,3], Yi Jiao (焦毅)[1,4], Ming-Yang Huang (黄明阳)[1,2], Shou-Yan Xu (许守彦)[1,2], Na Wang (王娜)[1,4], Sheng Wang (王生)[1,2; 1)]

[1] Institute of High Energy Physics, Chinese Academy of Sciences, Beijing 100049, China
[2] Dongguan Institute of Neutron Science (DINS), Dongguan 523808, China
[3] University of Chinese Academy of Sciences, Beijing 100049, China
[4] Key Laboratory of Particle Acceleration Physics and Technology, Institute of High Energy Physics, Chinese Academy of Sciences

**Abstract:** The Robust Conjugate Direction Search (RCDS) method is used to optimize the collimation system for the Rapid Cycling Synchrotron (RCS) of the China Spallation Neutron Source (CSNS). The parameters of secondary collimators are optimized for a better performance of the collimation system. To improve the efficiency of the optimization, the Objective Ring Beam Injection and Tracking (ORBIT) parallel module combined with MATLAB parallel computing is used, which can run multiple ORBIT instances simultaneously. This study presents a way to find an optimal parameter combination of the secondary collimators for a machine model in preparation for CSNS/RCS commissioning.
**Key words:** CSNS, RCDS, collimation system, optimization, parallel computing
**PACS:** 29.25.Dz, 29.27.-a, 41.85.Si

## 1 Introduction

The China Spallation Neutron Source (CSNS) is designed to provide a proton beam with beam power of 100 kW [1, 2]. The accelerator complex consists of an 80 MeV Linac and a 1.6 GeV Rapid Cycling Synchrotron (RCS) [3, 4]. The Linac accelerates the H$^-$ beam produced by an ion source. The RCS accelerates a proton beam which is converted from the H$^-$ beam by a stripping foil. In the RCS, the proton beam is accumulated through an anti-correlated painting scheme within 200 turns, and accelerated to 1.6 GeV in about 20000 turns [5, 6].

The RCS lattice has a four-fold structure with four straight sections designed for beam injection, transverse collimation, extraction, and RF systems, respectively. The main parameters of the RCS are listed in Table 1.

For the RCS, with beam energy ranging from 80 MeV to 1.6 GeV, the space charge forces are strong and have a large impact on beam dynamics. The emittance growth and halo generation induced by space charge could lead to unacceptably high beam loss [7, 8]. Considering the requirements for hands-on and safe maintenance of the machine, the average particle loss should be controlled to a low level of 1 W/m [9]. To meet this requirement, a two-stage collimation system was designed to localize the beam loss in the collimation section in the RCS [10, 11].

Table 1. The main parameters of CSNS/RCS

| Parameters | Symbol, unit | Value |
|---|---|---|
| Circumference | $C$, m | 227.92 |
| Injection energy | $E_{\text{inj}}$, MeV | 80 |
| Extraction energy | $E_{\text{ext}}$, GeV | 1.6 |
| Betatron tune (H/V) | $\upsilon_x/\upsilon_y$ | 4.86/4.78 |
| Accumulated particles per bunch | $N_p$, $\times 10^{12}$ | 7.8 |
| Harmonic number | $h$ | 2 |
| Repetition frequency | $f_0$, Hz | 25 |
| Accumulated and accelerated time per cycle | $t$, ms | ~20 |
| Transverse acceptance | $\varepsilon$, $\pi$mm·mrad | 540 |

The transverse collimation system consists of one primary collimator and four secondary collimators. The primary collimator has minimum acceptance in the ring, and thus is used to scatter the protons with large deviations from the beam center.

* Supported by National Natural Science Foundation of China (11475202, 11405187, 11205185) and Youth Innovation Promotion Association of Chinese Academy of Sciences (No. 2015009)
1) E-mail:wangs@ihep.ac.cn

The scattered protons are then absorbed by four secondary collimators located downstream of the primary collimator. The layout of the transverse collimation system and the optical parameters are shown in Fig. 1.

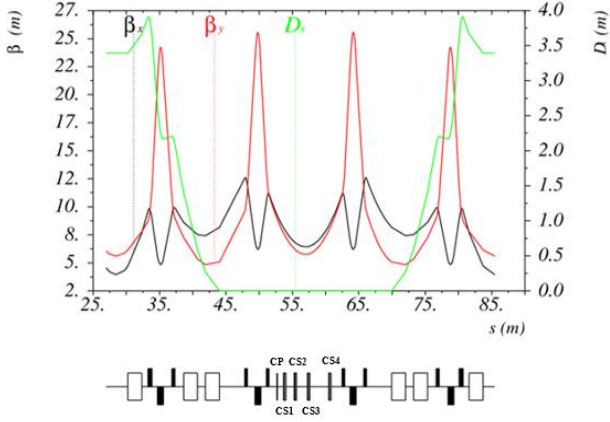

Fig. 1. (color online) Optical functions along a ring super period of the RCS, and the layout of the transverse collimation system. CP represents the primary collimator. CS1, CS2, CS3 and CS4 represent four secondary collimators in sequence, respectively).

In the RCS, the aperture of each secondary collimator can be varied by adjusting the positions of four movable blocks. In total there are sixteen tunable parameters. To achieve satisfactory machine performance, it is necessary to optimize these parameters, so as to absorb most of the undesirable protons by the collimation system and minimize the beam loss in other regions of the ring. Now the RCS is under construction, and the collimation efficiency with different sets of collimator parameters is evaluated with numerical simulation in this study. The collimation process in the presence of space charge is simulated with the Objective Ring Beam Injection and Tracking (ORBIT) code [12, 13].

In previous optimization studies of the collimation system, the acceptances of all secondary collimators were set to the same value, and several ORBIT simulations were performed with a rough grid scan method to select a set of parameters of collimators with relatively high collimation efficiency [11]. If there are $m$ values of the acceptance of each secondary collimator, then $m^4$ ORBIT instances need to be run for comparison. It is time consuming to find a suitable set of parameters through an overall comparison of all the possible combinations. Instead, in this study, we introduce an algorithm, the Robust Conjugated Direction Search (RCDS) method, in the optimization. This method is effective in optimizing a multi-variable objective online and it has both high tolerance to noise and high convergence speed [14]. It has been used for online optimization of machine performance when the objective function can be measured [14-16].

In Section 2, we will first discuss the modeling of the collimation system and concrete implementation of the ORBIT simulation and the RCDS algorithm. The optimization results will be presented in Section 3. A summary and discussions will be given in Section 4.

## 2 Physics analysis and modeling

To implement the application of the RCDS method in the optimization of the RCS collimation system, model parameters of an ORBIT instance to simulate the collimation process were first determined. Then a code was written in MATLAB to run the ORBIT instances automatically and to return the objective function, which is defined to describe the collimation efficiency, to the RCDS algorithm. In addition, to save the running time of the RCDS algorithm, parallel computing of the ORBIT instances was developed.

### 2.1 Physical variables

In this study, the acceptance of the primary collimator is fixed to 350 πmm·mrad all the time, and the secondary collimators are tuned to optimize the performance of the collimation system.

The structure of a secondary collimator is shown in Fig. 2. Each of the secondary collimators is composed of four movable copper blocks with thickness of 200 mm. Two of the blocks are in the vertical direction and the other two, downstream of the vertical blocks, are in the horizontal direction. All the blocks can be adjusted individually.

Each block has a circular surface based on the equation,

$$\frac{\left(x\cos(\frac{\pi\theta}{180°})+y\sin(\frac{\pi\theta}{180°})\pm c\right)^2}{a^2}+\frac{\left(-x\sin(\frac{\pi\theta}{180°})+y\cos(\frac{\pi\theta}{180°})\right)^2}{a^2}=1, \quad (1)$$

where $\theta = 0°$ or $90°$ corresponds to the horizontal or vertical direction of the block, $a$ is the radius, and $c$ is the distance between the beam center and the geometric center of the block, which can be changed from 34.8 mm to 68.8 mm determined by the mechanical design.

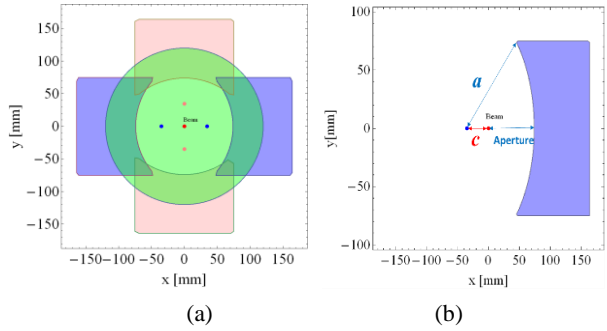

(a)          (b)

Fig. 2. (color online) The structure of a secondary collimator. (a) A second collimator has four blocks (pink: vertical blocks, blue: horizontal blocks) and has a cross section with the ring (green: vacuum chamber of the RCS). (b) The relationship between the parameter $c$ and the aperture of the acceptance of the collimator.

The parameter $c$ of each block is closely related to the acceptance of the collimator, so this parameter is selected to be the variable for the optimization. Considering the symmetry of beam distribution in simulations, parameters $c$ of the blocks on the same direction of each secondary collimator are same. Then there are eight variables to be tuned for four secondary collimators, i.e., ($c_1$, $c_2$, $c_3$, $c_4$, $c_5$, $c_6$, $c_7$, $c_8$), as shown in Fig. 3.

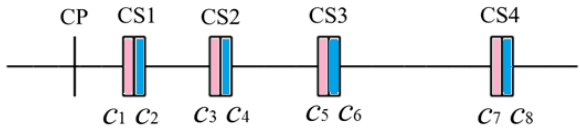

Fig. 3. (color online) The variables to be tuned in the optimization of the RCDS collimation system.

### 2.2 Objective function

The goal of the optimization of the collimation system is to localize the beam loss with a high shielding efficiency, and meanwhile, to have a low uncontrolled beam loss, a high cleaning efficiency of the system [17], a high collimation efficiency and a low beam loss around the RCS.

In our study, a single objective function is constructed to measure the performance of the collimation system, which is in the form of

$$f = -\eta_{\text{system}} \cdot R_{\varepsilon_{xm}} \cdot R_{\varepsilon_{ym}} \cdot R_{\varepsilon_{add}} \cdot R_{\varepsilon_{flag}}, \quad (2)$$

where a minus sign is added to form a minimization problem, and $\eta_{\text{system}}$ is the cleaning efficiency of the system,

$$\eta_{\text{system}} = \frac{N_{\text{incol}} + N_{\text{indrift}}}{N_{\text{loss}}}, \quad (3)$$

where $N_{\text{incol}}$ and $N_{\text{indrift}}$ are the numbers of particles absorbed by the collimators and the drifts between collimators, respectively, and $N_{\text{loss}}$ is the total number of particles lost in the RCS.

In Eq. (2), $R_{\varepsilon_{xm}}$ and $R_{\varepsilon_{ym}}$ are the weight factors satisfying the equations of

$$R_{\varepsilon_{xm}} = \begin{cases} 1, & \text{if } \varepsilon_{xm} \leq Ref\_collimator; \\ Ref\_collimator / \varepsilon_{xm}, & \text{if } \varepsilon_{xm} > Ref\_collimator. \end{cases} \quad (4)$$

$$R_{\varepsilon_{ym}} = \begin{cases} 1, & \text{if } \varepsilon_{ym} \leq Ref\_collimator; \\ Ref\_collimator / \varepsilon_{ym}, & \text{if } \varepsilon_{ym} > Ref\_collimator. \end{cases} \quad (5)$$

where the parameter $Ref\_collimator$ is set to 350 $\pi$mm·mrad to reflect the limitation from the primary collimator, and $\varepsilon_{xm}$ and $\varepsilon_{ym}$ are the maximum 99% horizontal and vertical emittances of the beam, respectively.

In Eq. (2), the weight factor $R_{\varepsilon_{add}}$ satisfies the equation of

$$R_{\varepsilon_{add}} = \begin{cases} 1, & \text{if } \varepsilon_{add} \leq Ref\_add; \\ Ref\_add / \varepsilon_{add}, & \text{if } \varepsilon_{add} > Ref\_add. \end{cases} \quad (6)$$

where the parameter $Ref\_add$ is set to 700 $\pi$mm·mrad for a further limitation on the beam emittance, and $\varepsilon_{add}$ is defined as

$$\varepsilon_{add} = \varepsilon_{xm} + \varepsilon_{ym}. \quad (7)$$

In Eq. (2), the weight factor $R_{\varepsilon_{flag}}$ is used to limit the shape of the emittance, and it satisfies the equation

$$R_{\varepsilon_{flag}} = \begin{cases} 1, & \text{if } \varepsilon_{flag} \leq Ref\_flag; \\ Ref\_flag / \varepsilon_{flag}, & \text{if } \varepsilon_{flag} > Ref\_flag. \end{cases} \quad (8)$$

where $Ref\_flag$ is set to 0.5 to make the horizontal emittance closer to the vertical emittance, and $\varepsilon_{flag}$ is defined as

$$\varepsilon_{flag} = \max(\frac{\varepsilon_{xm}}{\varepsilon_{ym}}, \frac{\varepsilon_{ym}}{\varepsilon_{xm}}) - 1. \quad (9)$$

### 2.3 Implementation of the simulation code

For each set of variable values of secondary collimators, ORBIT simulation is implemented to simulate the injection and acceleration process, from which the corresponding objective function is calculated. In the following, such a process is called an ORBIT instance.

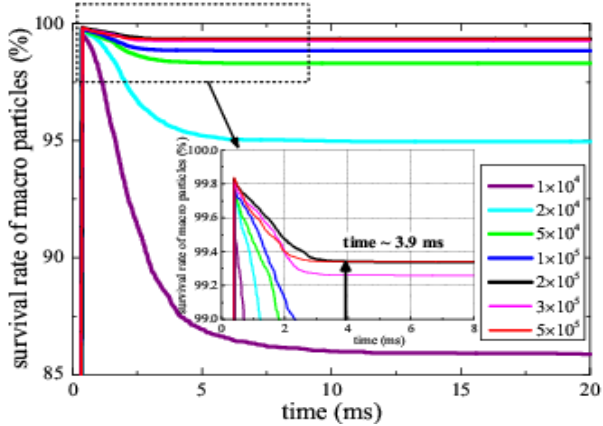

Fig. 4. (color online) Variation of the survival rate of macro particles with time for different particle numbers.

In the RCS, the number of protons in one bunch is $7.8 \times 10^{12}$. They are modeled as macro particles with a smaller number of $2 \times 10^5$ for an ORBIT instance. Simulations have been performed to compare the survival rates of macro particles for different numbers of particle, as shown in Fig. 4. This shows that with $2 \times 10^5$ macro particles, one can simultaneously achieve high computing precision and fast execution speed of the program. It also shows with such a number of macro particles, the beam loss mainly occurs within the first 3.9 ms, due to low energy, bunching process and the space charge effect.

So the beam is tracked for 2200 turns, which corresponds to a time of about 3.9 ms for accumulation and acceleration, instead of about 20000 turns, which takes about 20 ms for accumulation and acceleration, for enough stability of simulation results and a short CPU time. Values of the model parameters are listed in Table 2.

Table 2. The values of the model parameters

| Model parameters / unit | Value |
|---|---|
| number of real protons / $\times 10^{12}$ | 7.8 |
| number of macro particles / $\times 10^5$ | 2 |
| number of injection turns | 200 |
| number of acceleration turns | 2000 |
| beam tracking, turns / time | 2200 turns / 3.9 ms |

In an ORBIT instance, macro particles are generated and put into the injection simulation. The output of the injection simulation is then used as input for the acceleration simulation. Figure 5 shows the schematic of running an ORBIT instance and data stream between the injection and acceleration.

To accommodate the searching process of the algorithm in an automatic way, a code was first written in MATLAB to create and submit an ORBIT instance automatically. In addition, an identifier, an output document (named orbit_bye) was generated to signify the end of an instance, as shown in Fig. 5. Finally, the objective function was calculated from the output both of the injection and acceleration.

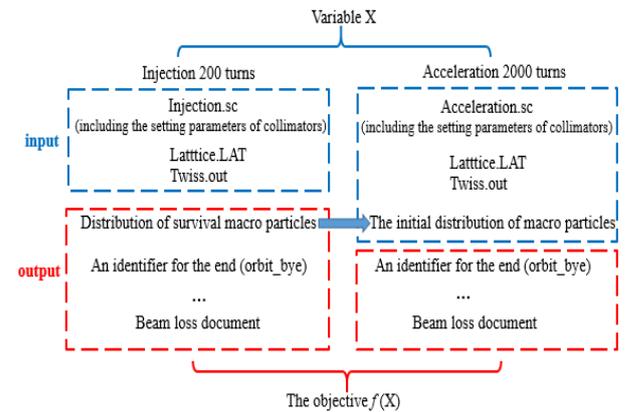

Fig. 5. (color online) Schematic of running an ORBIT instance and data stream between the injection and acceleration. The input files, Lattice.LAT and Twiss.out, are produced by the MAD program [18] and they contain the information on the RCS lattice.

## 2.4 Parallel computing of ORBIT instances

The ORBIT code can run with the Message Passing Interface (MPI) parallel computing on a LINUX parallel computer. Here a cluster in a Gigabit network environment at IHEP was used. Each node of the cluster has 2 CPUs, and each CPU has 6 cores. Limited by available network bandwidth and information exchange between nodes (e.g., the transverse space charge calculation [12]), the communication loss between nodes is not ignorable. So when the cores belong to different nodes, an increase of the number of cores would not lead to a proportionate decrease in the CPU time of running an ORBIT instance.

To make better use of nodes in the cluster and speed up computation of the RCDS method, 12 cores inside the same node were used to run an instance with the MPI parallel computing, and meanwhile a function was programmed within the RCDS algorithm to run several instances on different nodes simultaneously with the MATLAB parallel computing toolbox (PCT) [19].

In the RCDS method, a series of one-dimensional searches are run to implement the conjugate direction search. In the robust line optimizer, the algorithm uses golden section extrapolation to determine the bounds. The original RCDS method takes the evaluations in sequence as needed.

For a certain direction, the starting point, the step size and the range for the variables can be determined before the line optimizer is executed. So the candidate values of variables are calculated to give out a variables vector set ($X_1$, $X_2$, …, $X_n$), and the corresponding ORBIT instances are created. At this point, the function we defined in MATLAB is called to distribute these ORBIT instances over $n$ nodes individually with PCT, and then run each ORBIT instance inside a separate node with MPI. Finally, the corresponding objective values across all the separate nodes are obtained to form a library. Figure 6 shows the structure of the function. A candidate function library, $\{(X_1, f_1), (X_2, f_2), …, (X_n, f_n)\}$, for the present direction is set up. So the following line optimizer in the direction can search the candidate function library and get the matching line of variables among the candidates directly, which further enhances the simulation efficiency.

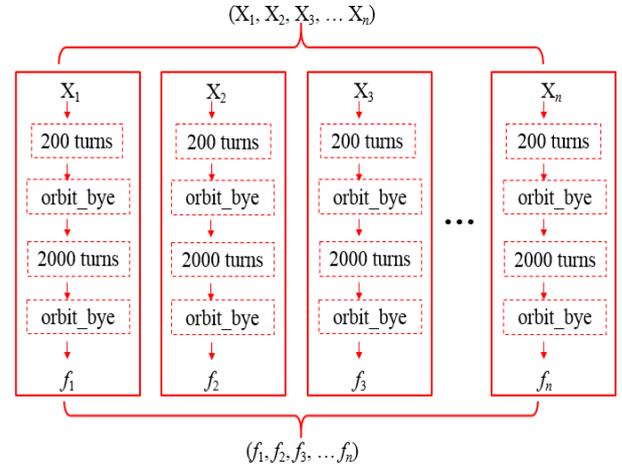

Fig. 6. The structure of the function.

In this method, a group of candidate evaluations of the objective are obtained in a period of time during which the original method can get only one evaluation. Thus, by this method, the CPU time of running several ORBIT instances decreases in proportion to the number of nodes with the same cores being used on each node.

## 3 Optimization result

As the beam distribution for the input of the injection process of an ORBIT instance is given by the Twiss parameters, there are random factors during the generation process. The beam distribution is also determined by the size of the injection beam from the Linac and the injection painting scheme in the actual operation, so the beam loss will also be affected. Therefore, in order to confirm the optimization of the performance of the collimation system, we fixed the particle distribution for the input of the acceleration process. A realistic distribution of macro particles was obtained with the acceptances of secondary collimators being set to 500 $\pi$mm·mrad, as shown in Fig. 7. The horizontal 99% emittance of the beam distribution is 193 $\pi$mm·mrad and the vertical is 219 $\pi$mm·mrad.

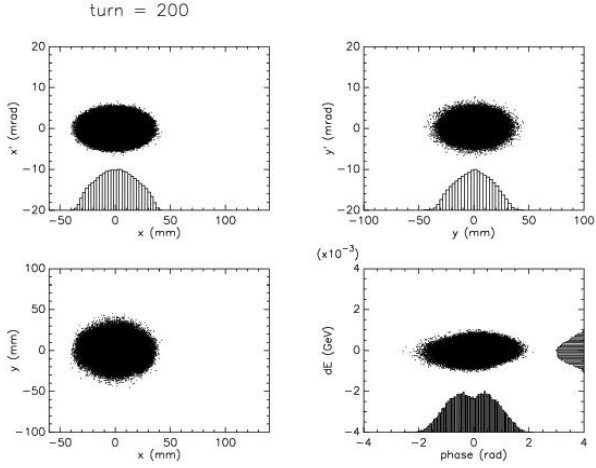

Fig. 7. The beam distribution used as input for the acceleration.

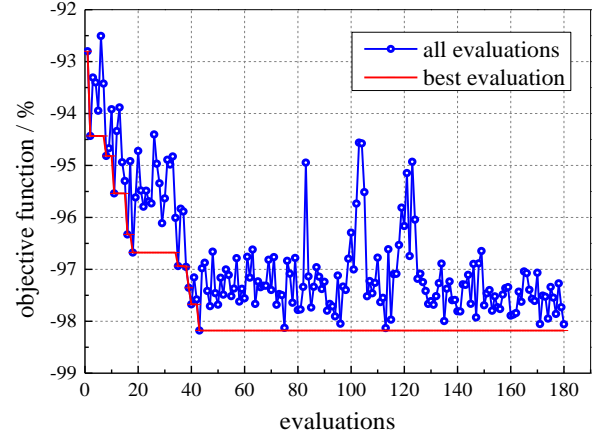

Fig. 8. History of all evaluations and the best evaluations during the optimization of the collimation system with the RCDS method.

In the following, the performance of the collimation system during the acceleration process is presented and analyzed. By running an instance with the beam being accelerated for 2000 turns repeatedly, the noise level of the cleaning efficiency was calculated to be 0.7%, and this value was used as the noise of the objective during the optimization. The initial acceptances of secondary collimators were set to 420 $\pi$mm·mrad. Their ranges were tuned from 370 $\pi$mm·mrad to 500 $\pi$mm·mrad due to the acceptance of the primary collimator and the transverse acceptance of the ring. Based on the optical functions, as shown in Fig. 1, the initial values of the variables were calculated and listed in Table 3.

Table 3. The initial values of the variables.

| variables | $c_1$ | $c_2$ | $c_3$ | $c_4$ | $c_5$ | $c_6$ | $c_7$ | $c_8$ |
|---|---|---|---|---|---|---|---|---|
| value / mm | 52.8 | 51.6 | 57.9 | 55.7 | 59.5 | 56.6 | 50.6 | 48.2 |

Having configured the parameters of RCDS and given the initial values of the variables, the simulation was performed to optimize the collimation system for the RCS. Figure 8 shows the objective function for all trial evaluations and the best evaluations. The objective was optimized from −92.8% to −98.2% with the product of the weight factors set to 1. With eight variables, the objective was optimized automatically within 180 evaluations. The simulation time was lower than the time of running one ORBIT instance 180 times.

Table 4 shows a comparison of the parameters reflecting the performance of the collimation system between the initial state and the optimal result. The cleaning efficiency, $\eta_{system}$, was optimized to 98.2%. The total beam loss along the ring was acceptable for shielding, although it was a little higher than that of the initial state. The uncontrolled beam loss of $1.7 \times 10^{-4}$ of the total beam was lower. Considering even larger beam loss might be caused by various kinds of errors in the actual conditions, it is more important to reduce the uncontrolled beam loss. The optimal values of the variables and the beam losses in the secondary collimators as a percentage of the total beam loss are shown in Table 5.

Table 4. Comparison of parameters reflecting the performance of the collimation system between the initial state and the optimal result. The parameter $\eta_{system}$ is the cleaning efficiency given by Eq. (3), $\eta_{co}$ is the collimation efficiency of collimators, $\lambda_{un}$ is the uncontrolled beam loss of total beam outside the collimation section during the acceleration, $\lambda_{total}$ represents the total beam loss as a percentage of total beam along the ring during acceleration, and $\varepsilon_x$ ($\varepsilon_y$) is the 99% horizontal (vertical) emittance of the beam.

| parameters | $\eta_{system}$ / % | $\lambda_{un}$ / $10^{-4}$ | $\eta_{co}$ / % | $\lambda_{total}$ / % | $\varepsilon_x$ /$\pi$mm·mrad | $\varepsilon_y$ /$\pi$mm·mrad |
|---|---|---|---|---|---|---|
| initial state | 92.8 | 4.9 | 91.9 | 0.7 | 193 | 219 |
| optimal result | 98.2 | 1.7 | 96.3 | 0.9 | 193 | 215 |

Table 5. The optimal values of the variables and the beam losses in the secondary collimators as a percentage of total beam loss. The parameter $\lambda_{co}$ is the beam loss in the secondary collimator as a percentage of total beam loss during acceleration.

| Element | variables | value / mm | Acceptance / π mm·mrad | $\lambda_{co}$ / % |
|---------|-----------|------------|------------------------|--------------------|
| CS1 | $c_1$ | 56.1 | 371 | 29.58 |
|     | $c_2$ | 48.3 | 470 |  |
| CS2 | $c_3$ | 60.9 | 372 | 58.83 |
|     | $c_4$ | 55.8 | 419 |  |
| CS3 | $c_5$ | 59.5 | 420 | 5.08 |
|     | $c_6$ | 56.6 | 420 |  |
| CS4 | $c_7$ | 50.6 | 420 | 2.65 |
|     | $c_8$ | 48.2 | 420 |  |

## 4  Summary and Discussions

In this paper, we have implemented the RCDS method to optimize the collimation system of CSNS/RCS. The uncontrolled beam loss of the total beam during the acceleration can be reduced to $1.7 \times 10^{-4}$, which is lower than that obtained by previous optimization [11]. As a result, an approach was established to efficiently give an optimal parameter combination of the secondary collimators for the present machine model.

From the optimized result, the beam loss mostly occurs in the first two collimators. This indicates the possibility of using fewer secondary collimators. But, it may also originate from the intrinsic characteristic of using the unit vectors as the initial direction set for the algorithm, which should be verified with more studies in the future.

## 5  Acknowledgment

The authors thank X. B. Huang for a lot of helpful discussions.